\documentclass[usenatbib]{mn2e}
\usepackage[utf8]{inputenc}
\usepackage{natbib,epsfig,amsmath,cite,amssymb,graphicx,mystyle}
\usepackage{mathtools}
\usepackage{array, xcolor, caption, multirow}
\usepackage[colorlinks=true,linkcolor=blue,citecolor=blue]{hyperref}

\title[Gravitational-wave background from stalling massive black-hole binaries]{The nightmare scenario: measuring the stochastic  gravitational-wave background from stalling massive black-hole binaries with pulsar-timing arrays}

\author[Dvorkin \& Barausse]{Irina Dvorkin$^{1,2}$\thanks{E-mail: dvorkin@iap.fr} \& Enrico Barausse$^{1}$\thanks{E-mail: barausse@iap.fr}\\
$^{1}$ Institut d'Astrophysique de Paris, Sorbonne Universit\'{e}s, UPMC Univ Paris 6 \& CNRS, UMR 7095, 98 bis bd Arago,\\ F-75014 Paris, France\\
$^{2}$ Institut Lagrange de Paris (ILP), Sorbonne Universit\'{e}s, 98 bis bd Arago, F-75014 Paris, France}

\begin{document}

\pagerange{\pageref{firstpage}--\pageref{lastpage}} \pubyear{2017}
\maketitle
\label{firstpage}

\begin{abstract}
Massive black-hole binaries, formed when galaxies merge, are among the primary sources of gravitational waves targeted by ongoing
Pulsar Timing Array (PTA) experiments and the upcoming space-based LISA interferometer. However, their formation and merger rates are still
highly uncertain. Recent upper limits on the stochastic gravitational-wave background obtained by PTAs
are starting to be in marginal tension with theoretical models for the pairing and orbital evolution of these systems.
This tension can be resolved by assuming that these binaries are more eccentric
or interact more strongly with the environment (gas and stars) than expected, or by accounting
for possible selection biases in the construction of the theoretical models. However, another (pessimistic) possibility
is that these binaries do not merge at all, but stall at large ($\sim$ pc) separations.
We explore this extreme scenario by using a semi-analytic galaxy formation model
including massive black holes (isolated and in binaries), and show that future generations of PTAs
will detect the stochastic gravitational-wave background from the massive black-hole binary population
within $10-15$ years of observations, even in the ``nightmare scenario'' in which all binaries stall at
the hardening radius. Moreover, we argue that this scenario is too pessimistic, because our
model predicts the existence of a sub-population of binaries with small mass ratios ($q \lesssim 10^{-3}$) that
should merge within a Hubble time simply as a result of gravitational-wave emission. This sub-population
will be observable  with large signal-to-noise ratios  by future PTAs thanks to next-generation radio telescopes such
as SKA or FAST, and possibly by LISA.
\end{abstract}


\section{Introduction}

Massive black holes (MBHs) with masses in the range $\sim 10^6-10^{9}M_{\odot}$ are ubiquitous in the nuclei of nearby and distant galaxies \citep{1995ARA&A..33..581K}. 
In the accepted framework of hierarchical structure formation, massive galaxies are formed by continuous accretion of dark matter and gas from cosmic filaments, and by (minor and major) galaxy mergers. 
The latter process is expected to lead to the formation of a population of MBH binaries in the nuclei of post-merger galaxies~\citep{1980Natur.287..307B}. 
If these binaries are at sufficiently close separations, they efficiently emit gravitational waves (GWs), which may be observable
by ongoing Pulsar Timing Array (PTA) experiments~\citep{1983ApJ...265L..39H} 
for total binary masses of $\sim10^8$-$10^{10} M_\odot$ and separations of hundreds to thousands of gravitational radii.
These experiments include the European Pulsar Timing Array (EPTA;~\citealt{epta}), 
the Parkes Pulsar Timing Array (PPTA;~\citealt{ppta}) and the North American Nanohertz Observatory for Gravitational Waves (NANOGrav;~\citealt{nanograv}), 
joining together in the International Pulsar Timing Array (IPTA;~\citealt{ipta}).
Moreover, MBH binaries with total masses $\sim10^4$-$10^{7} M_\odot$ will also
be observable in the late inspiral, merger and ringdown phases
 by the upcoming space-borne Laser Interferometer Space Antenna (LISA;~\citealt{2017arXiv170200786A,2016PhRvD..93b4003K}). More precisely, both LISA and PTAs will not only target the GWs from individual resolved sources, but also the stochastic background resulting from the superposition
of the GWs produced by all the unresolved sources that exist in the universe.
Still, although LISA is expected to detect e.g. a significant stochastic background of Galactic white-dwarf binaries~\citep{2017arXiv170200786A}, unresolved MBH binaries are expected to be relatively rare in the LISA frequency range (i.e.
LISA will detect the majority of the MBH mergers in the Universe, or even all of them depending on the astrophysical model, c.f.~\citealt{2016PhRvD..93b4003K}).
Conversely, PTAs are expected to first detect the stochastic GW background from MBH binaries, though they are sensitive also to signals from individual loud sources \citep{2015MNRAS.451.2417R}.

Indeed, PTA upper limits on the stochastic GW background from MBH binaries have steadily improved over the past few years, and they have recently started
being in marginal tension with the predictions of theoretical models~\citep{2015MNRAS.453.2576L,2016ApJ...821...13A,2015Sci...349.1522S}. This is not surprising in itself,  because
when two galaxies merge, the MBHs are expected to be deposited in the outskirts of
the newly formed galaxy, at separations that could be as large as $\sim$ kpc.
Early on, dynamical friction from the stellar and gas background is probably
very efficient at driving the MBHs towards the galactic centre (since 
most MBHs will be still surrounded by a relic stellar cluster inherited from 
their previous host galaxy). However,  when the MBHs form a bound binary, dynamical
friction becomes inefficient, because the  binary's orbital velocity exceeds the typical velocity of the stars.
The subsequent shrinking of the binary is then assured by three-body interactions with stars. Indeed, stars with
angular momentum in an appropriate region of parameter space (the ``loss cone'') will interact strongly with the
binary and typically extract energy from it. As a result, the binary will shrink~\citep{1996NewA....1...35Q}, while
the stars will be scattered away and possibly ejected from the galaxy as hypervelocity stars~\citep{2006ApJ...651..392S}.
After a phase of fast shrinking, 
% the binary starts hardening at a constant rate when its separation reaches 
the binary starts hardening at a rate $\dot{a}\propto a^2$ when its separation $a$ reaches the hardening radius $a_{\rm h}$ ($a_{\rm h}\sim$ pc for total masses of $\sim 10^8 M_\odot$, c.f. Equation \ref{eq:ahard} below). This will eventually 
 deplete the ``loss cone'' in the phase space of the surrounding
 stars, which will result in three-body interactions also becoming inefficient.
 While the loss 
 cone will be replenished naturally by the scattering between stars on the relaxation timescale, 
 this exceeds the Hubble time for galaxies hosting MBHs with masses above $\sim10^9 M_\odot$. 
 Since GW emission does not become efficient enough to drive the binary
 to merger within a Hubble time until the separation becomes of the order of $a_{\rm gw}\sim 10^{-2}$ pc (for total masses of $\sim 10^8 M_\odot$, c.f. Equation \ref{eq:agw} below),
 binaries with large masses ($\gtrsim 10^8 - 10^9 M_\odot$)  may therefore stall at separations $a\lesssim a_{\rm h}\sim$ pc. This is known as the ``last-parsec problem''~\citep[c.f.][]{2001ApJ...563...34M,2005LRR.....8....8M,2011ApJ...732L..26P,2014SSRv..183..189C}.
 
A way around this problem is provided by processes that could help replenish the loss cone, e.g.
galaxy rotation~\citep{2015arXiv150506203H} or a tri-axiality in the galactic gravitational 
potential~\citep{yu,Khan2011,2014CQGra..31x4002V,lastPc2,2015ApJ...810...49V,lastPc2,2015MNRAS.454L..66S}, induced for instance by mergers.
This replenishment would make stellar scattering efficient again at driving the orbital evolution
down to the separation $a_{\rm gw}$ needed for the binary to merge as a result of GW emission. Other possibilities
to overcome the last-parsec problem are the presence of a gaseous disc, which would result in planetary-like migration of the MBHs towards the centre~(\citealt{bence}; see 
however \citealt{2009MNRAS.398.1392L} for complications arising in this scenario);
or the interaction with a third incoming MBH coming from a subsequent galaxy merger, which would trigger the coalescence of the
binary via the combined action of Kozai-Lidov resonances and GW emission~\citep{2007MNRAS.377..957H,2016MNRAS.461.4419B}.

Nevertheless, while there is a consensus that the last-parsec problem will be somehow solved, the exact mechanism by which this would happen is still debated.
As a matter of fact, the aforementioned PTA limits on the stochastic GW background are starting to probe (and even in some cases to be in marginal tension
with) models for MBH binary formation and evolution that assume that all binaries merge efficiently under the effect of GW emission alone~\citep{2015MNRAS.453.2576L,2016ApJ...821...13A,2015Sci...349.1522S}.
Several ways to explain the PTA limits have been proposed. 
MBH binaries are normally assumed to be almost circular when they enter the PTA band, but they could have a significant non-zero eccentricity~\citep{2017PhRvL.118r1102T}, which
could be left over for instance from triple MBH Kozai-Lidov oscillations and which 
 would move at least part of the radiated power outside the PTA band.
Binaries may also interact more strongly with the environment (gas and stars) than expected~\citep{2011MNRAS.411.1467K,2015PhRvD..91h4055S,2014MNRAS.442...56R,vallisneri,2017arXiv170202180K}, and these interactions
may still be important at the separations that are most relevant for PTAs ($a\lesssim 10^{-2}$ pc). As a result, a binary's orbital energy would
be lost at least partly to the environment (as heating of the gas or increase in the stars' velocities) as the system inspirals, rather than be emitted in GWs alone.
Yet another possibility is that the theoretical predictions for the {\it number} of MBH binaries are off, since they are 
produced with models calibrated to the MBH scaling relations, which may be biased-high due to selection effects (\citealt{2016MNRAS.463L...6S}; see also \citealt{shankar,2017MNRAS.tmp....6S,2017MNRAS.468.4782B}).
However, the simplest and most pessimistic possibility is that the last-parsec problem may {\it not} be solved after all, and that MBHs may stall at separations $\sim a_{\rm h}$ or even larger.

In this paper we explore this ``nightmare-scenario'' by
using a comprehensive semi-analytic galaxy-formation model~\citep{2012MNRAS.423.2533B},
which includes MBHs (in isolation and in binaries) and their co-evolution with their host galaxies.
We  show that while
the stochastic GW background
predicted within this pessimistic scenario is way outside the reach of
current experiments, it will still be detectable  within $10-15$ years of observations
by future PTA experiments, thanks to next-generation radio telescopes with large collecting areas such as the Square Kilometre Array radio-telescope~\citep[SKA;][]{ska} or the 
Five hundred meter Aperture Spherical Telescope~\citep[FAST;][]{2011IJMPD..20..989N}. Moreover,
we show that even if we insist that all binaries should stall at the hardening radius $a_{\rm h}$, our semi-analytic model predicts
the existence of a non-negligible sub-population of binaries with small mass ratios ($q \lesssim 10^{-3}$). These binaries would
coalesce in less than a Hubble time if initially placed at a separation $a_{\rm h}$. This would significantly increase
the expected stochastic background signal, which should be observable with very high signal-to-noise ratios by SKA- or FAST-based PTAs.
We also show that the formation of this sub-population of binaries is not an artifact of the simplified prescriptions used in the
semi-analytic model to account for the orbital evolution of merging galaxies and MBH binaries. Moreover, we 
show that this subpopulation may also give rise to a few events detectable by the LISA mission, in the form of intermediate mass-ratio inspirals \citep[IMRIs;][]{imris1,imris2,imris3}, if MBHs
form from ``light'' seeds ($M_{\rm seed}\sim200\, M_\odot$), e.g. the remnants of popIII stars.

The structure of this paper is as follows. In Section \ref{sec:gwb} we derive a general expression for the stochastic 
GW background from a population of stalling binaries. In Section \ref{sec:formation} we present our semi-analytic galaxy formation  model 
and show that it predicts the existence of a sub-population of binaries that merge within a Hubble time 
from their hardening radius. We also outline in more detail the model for stalling MBH binaries that we utilise in this paper. 
The stochastic GW background from stalling and merging MBH binaries and its detection 
prospects are presented in Section \ref{sec:res}. We lay out our conclusions in Section \ref{sec:dis}. All cosmological parameters are taken from \citet{2016A&A...594A..13P}.

 \section{Gravitational-wave background}
 \label{sec:gwb}

The stochastic background of GWs with energy density $\rho_{\rm gw}$ can be characterised by the dimensionless parameter
\begin{equation}
\Omega_{\rm gw}(f)=\frac{1}{\rho_c c^2}\frac{d\rho_{\rm gw}}{d\ln f}
\end{equation}
where $\rho_c=3H_0^2/8\pi G$ is the critical mass density of the Universe, $H_0\approx 68$ km/s/Mpc is the Hubble constant and $f$ is the frequency measured in the detector frame, which is related to the frequency $f_s$ in the source frame 
by $f_s =f(1+z)$, where $z$ is the redshift. 
Let us consider a population of binary systems with comoving number density $n(M_c,z)$, where each binary is 
characterised by the masses of the components $m_1,m_2$, the chirp mass $M_c=(m_1 m_2)^{3/5}/(m_1+m_2)^{1/5}$, 
and the separation $a$ (we will assume circular orbits throughout this paper). The total energy density resulting from the emission of GWs by these sources 
is \citep{2001astro.ph..8028P,2008MNRAS.390..192S,2011PhRvD..84h4004R}:\footnote{Note that another
quantity widely used to characterise a GW stochastic background is the characteristic strain $h_{\rm c}(f)$, which is related to $\Omega_{\rm gw}(f)$
by $\Omega_{\rm gw}(f)={2\pi^2} \left[f h_c(f)\right]^2/{(3H_0^2)}$.
}
\begin{equation}\label{eq:Omega}
 \Omega_{\rm gw}(f)=\frac{f}{\rho_c c^2}\int dM_c dz \frac{d^2n}{dM_c dz}\frac{dE}{d f}\:.
\end{equation}
If a binary  overcomes the last-parsec problem  and the merger takes place at redshift $z$
 on timescales much shorter than the Hubble time, 
the observed spectrum $dE/df$ is related to the emitted spectrum $dE_s/df_s$ via
\begin{equation}\label{eq:dEdf1}
 \frac{dE}{df}(f)=\frac{dE_s}{df_s}\left(f_s\right)\,
\end{equation}
with 
\begin{equation}\label{eq:dEdf2}
 \frac{dE_s}{d\ln f_s}=\frac{(G\pi)^{2/3}}{3}M_c^{5/3}f_s^{2/3}\:.
\end{equation}
If all binaries merge efficiently, Equations \ref{eq:Omega}, \ref{eq:dEdf1} and  \ref{eq:dEdf2}
therefore give the power law
\begin{equation}
 \Omega_{\rm gw}(f)=\frac{(G\pi)^{2/3}}{3} \frac{f^{2/3}}{\rho_c c^2}\int dM_c dz \frac{d^2n}{dM_c dz}\frac{M_c^{5/3}}{(1+z)^{1/3}}\,.
 \label{eq:back_merging}
\end{equation}
 In practice, this power law can be suppressed at low frequencies, depending 
 on whether environmental effects (i.e. interactions with stars and gas) 
 are still important at the separations $a\lesssim 10^{-2}$ pc that are most important for PTAs.
 On the other hand, at high frequencies the signal is dominated by high-mass binaries ($M_c\gtrsim 10^8 M_{\odot}$ for $f \gtrsim \mbox{a few}\times 10^{-8}$ Hz, c.f. \citealt{2008MNRAS.390..192S}, hereafter SVC08). 
 Since these systems are intrinsically rare, a given realisation of the Universe may contain (on average) less than one such source close enough to contribute significantly to an observed frequency bin. 
 SVC08 showed that above $\mbox{a few}\times 10^{-8}$ Hz, this small number statistics effect causes 
simulated realisations of $\Omega_{\rm gw}(f)$ to display excess power (relative to Equation~\ref{eq:back_merging}) in (few) frequency bins, and lower power in the remaining ones. As a result, the slope of
$\Omega_{\rm gw}(f)$ at $f \gtrsim \mbox{a few}\times 10^{-8}$ Hz is flatter than predicted by Equation \ref{eq:back_merging}, and may even become zero or change sign at high frequencies~(SVC08; c.f. also black lines in Figure \ref{fig:stalling_omega} later on). 

To model binaries that do {\it not} merge in a Hubble time but stall at a separation $a_{\rm stall}$, we adopt two complementary approaches (in practice, as we will see, the results are close). In the first, we
simply use Equations~\ref{eq:Omega}, \ref{eq:dEdf1} and \ref{eq:dEdf2}, but cut off the spectrum given by 
Equation \ref{eq:dEdf2} outside the small frequency interval $[f_{\rm stall},f_{\rm stall}+\Delta f]$. Here, 
 $f_{\rm stall}=2f_0$, where $f_0$ is the orbital frequency $(2\pi f_0)^2=G(m_1+m_2)/a_{\rm stall}^3$,
while the (small) frequency shift $\Delta f$ is computed by evolving the binary under GW emission, from its formation redshift (at which
the separation is $a_{\rm stall}$) to the present time.

This approach relies on Equations~\ref{eq:dEdf1} and  \ref{eq:dEdf2}, which assume that GW emission happens on
timescales much shorter than the Hubble time. Since this is only approximately valid for stalling binaries,
we also do the calculation by assuming that the binary emits at the stalling frequency $f_{\rm stall}=2f_0$ \emph{at all times after formation}.
This assumption, while approximate in a different way (as in reality the frequency will slowly shift due to GW emission), allows us
to account for the changing redshift of the universe, i.e. 
 each stalling binary contributes to a range of detector-frame frequencies $f<f_{\rm stall}$. In more detail, since for
an unevolving stalling binary $f=f_{\rm stall}/(1+z)$ and thus $d\ln f = -d\ln (1+z)$, the detector-frame spectrum at frequency $f$ can be expressed as
\begin{equation}
\label{eq:stalling_spectrum}
 \frac{dE}{d\ln f}(f)=\frac{dE_s}{dt_s}(f_{\rm stall})\left\vert \frac{dt_s}{dz}(\bar{z})\right\vert\:,
\end{equation} 
where the emitted power is given by the quadrupole formula
\begin{equation}
 \frac{dE_s}{dt_s}(f_{\rm stall})=\frac{32c^5}{5G}\left(\frac{GM_c}{c^3}\pi f_{\rm stall} \right)^{10/3},
 \label{eq:dEdt}
\end{equation}
while 
\[
     \left\vert \frac{dt_s}{dz}(\bar{z})\right\vert=
\begin{dcases}
   \frac{1}{H_0\sqrt{\Delta(\bar{z})}(1+\bar{z})}&\text{if } \bar{z}\geq 0\\       0,         &\text{otherwise}
\end{dcases}
\]
with $\bar{z}\equiv f_{\rm stall}/f-1$ and $\Delta(\bar{z})\equiv\Omega_m (1+\bar{z})^3+\Omega_\Lambda$ (where $\Omega_m\approx0.3$ and $\Omega_\Lambda\approx0.7$ are the density parameters of matter and cosmological constant).
Note that as a result of the Heaviside function in the definition above, ${dE}/{d\ln f}(f)=0$ if $f>f_{\rm stall}$, i.e. stalling binaries
only emit at redshifted frequencies $f<f_{\rm stall}$. 
 The energy density due to a population of stalling binaries is then simply given by integrating over all sources, i.e.
\begin{equation}
 \Omega_{\rm gw}(f)=\frac{1}{\rho_c c^2}\int dM_c dz \frac{d^2n}{dM_c dz} \frac{dE_s}{dt_s}\left\vert \frac{dt_s}{dz}\right\vert
 \label{eq:back_stalling}\,.
\end{equation}

\section{The model}
\label{sec:formation}

\subsection{Semi-analytic galaxy formation model}

We follow the mergers of MBHs by
a state-of-the-art semi-analytic galaxy formation model introduced in \citet{2012MNRAS.423.2533B}, with later 
updates to improve the spin of evolution of MBHs~\citep{2014ApJ...794..104S} and to include nuclear star clusters in the centre of galaxies~\citep{2015ApJ...812...72A,2015ApJ...806L...8A}.
 This model accounts for the cosmological evolution and merger history of galaxies inside dark-matter halos, which are produced with Press-Schechter algorithms  calibrated to match the results of N-body simulations~\citep{Press74,Parki08}. Galaxies form from the cooling of a ``hot'' unprocessed intergalactic medium shock heated to
 the halo's virial temperature, or by cold accretion flows in low-mass and high-redshift systems~\citep{2006MNRAS.368....2D,2006MNRAS.370.1651C}. Once cooled or accreted to the halo's centre, the
 gas settles on a disc-like geometry by conservation of angular momentum, and eventually undergoes star formation. Bulges  form as a result of either major galactic mergers or bar instabilities, which both destroy the stellar and gaseous discs
 and typically trigger bursts of star formation as the gas is funneled towards the central ($\sim$ kpc) region of the galaxy.
 Whenever star formation takes place (in the bulge or in the disc) we account for the feedback of supernova explosions, which remove gas and tend to quench star formation, preferentially in low-mass systems.

The model also accounts for the presence of MBHs, which grow from high-redshift seeds
with mass of either  $M_{\rm seed}\sim200\, M_\odot$ \citep[``light seeds'', arising e.g. from the remnants of popIII stars; ][]{2001ApJ...551L..27M}
or $M_{\rm seed}\sim 10^5\, M_\odot$ \citep[``heavy seeds'', resulting for instance from protogalactic disc instabilities; ][]{2008MNRAS.383.1079V}.
These seed black holes are assumed to only form at $z>15$, with halo occupation fractions that
depend on their exact formation mechanism \citep[c.f.][for details]{2016PhRvD..93b4003K}. Note however that at $z\sim 0$ 
the predictions of our model are essentially independent of the seed model, at least in large systems, because accretion 
and mergers wash out the effect of the initial conditions as time progresses. In this paper we will see
an example of this fact, already noted e.g. in \citet{2012MNRAS.423.2533B,2014ApJ...794..104S,2015ApJ...812...72A,2015ApJ...806L...8A}.\footnote{Note however that the merger rate for
MBHs with mass between $10^4$ and $10^7 M_\odot$ -- i.e. the ones that are targeted by the LISA GW detector -- depend
more strongly on the seed model, c.f. \citet{2016PhRvD..93b4003K}.} 

The model assumes that MBHs accrete gas from a nuclear reservoir of cold gas, whose
 feeding correlates linearly with bulge star formation~\citep{2004ApJ...600..580G,2014ApJ...782...69L}. Since the latter
 takes place in our model following major mergers and disc instabilities, our MBHs undergo long periods of quiescent activity   occasionally interrupted by active quasar periods. The feedback of MBH activity
 on the surrounding gas (``AGN feedback'') is accounted for in both phases 
 (radio-mode and quasar feedback), 
 and quenches star formation in preferably high-mass systems.

Galaxy and black-hole mergers are modelled by starting from the halo merger history.
Whenever two halos coalesce in the extended Press-Schechter merger tree, we assume that the smaller one (together with the galaxy it hosts) survives as a subhalo/satellite galaxy inside the more massive host halo. We then account for the slow infall of this satellite to the centre of the host halo by dynamical friction, by using the expressions of \citet{boylankolchin}. Note that
these expressions are calibrated against N-body simulations and account for the dynamical friction due to both Dark Matter and baryons \citep[c.f. discussion in Section 2.3 of][]{boylankolchin}.\footnote{A subtle point in the implementation
of the dynamical friction timescale in a Dark-Matter merger tree is given by the treatment of the coalescence of 
halos (each of which will in general contain its own collection of subhalos). As in \citet{2012MNRAS.423.2533B}, if the coalescence has
(Dark-Matter) mass ratio $>0.3$, we re-initialize the dynamical friction times of all the subhalos. Otherwise, we reinitialize the
dynamical friction times of the subhalos of the ``satellite halo'', but keep those of the subhalos of the ``host'' unchanged.
This  corresponds to an intuitive scenario where the incoming halo perturbs and randomizes
the orbits of the host's subhalos, provided that it has a sufficiently large mass. We refer to \citet{2012MNRAS.423.2533B} for a more exhaustive
description of this and other details of the implementation.}
 During this slow infall, the outer regions of the satellite are tidally stripped, and the whole satellite
also undergoes tidal evaporation due to the fast-varying tidal field that it experiences near the periastron of its trajectory. We include both effects by modelling them after~\citet{2003MNRAS.341..434T}.
After this dynamical friction driven migration, the satellite galaxy finally reaches the centre of the host halo and
merges with the central galaxy. At this stage, the satellite MBH is expected to be 
left wandering in the outskirts of the newly formed galaxy and to fall towards its centre by
dynamical friction against the distribution of gas and stars. This process is normally thought to be quite efficient at shrinking the satellite MBH orbit
until it forms a bound binary with the central MBH (i.e. until the binary's orbital velocity exceeds the typical velocity of the stars). Indeed, typically the satellite MBH will still be surrounded by at least the inner regions of the satellite galaxy, which increase its effective mass and thus the efficiency of dynamical friction.
Nevertheless, we still account for this evolutionary phase in our model, in the case of systems with small mass ratios (i.e. satellite MBHs much smaller than the central MBH), 
for which the satellite black hole might be stripped of all its galaxy quite early on. 

As already mentioned in the Introduction, once a bound MBH binary is formed, dynamical friction becomes inefficient at driving the binary's evolution any further, but
three-body interactions of the binary with stars become important. 
These interactions tend to transfer energy from the binary, whose orbit shrinks, to the stars, which may even get ejected from the galaxy
as hypervelocity stars~\citep{2006ApJ...651..392S}. In more detail, after an initial fast shrinking of the orbit, the binary will harden at a constant rate after reaching the hardening radius~\citep{2006ApJ...648..976M}:
\begin{equation}
 a_{\rm h}=11\left(\frac{m_1+m_2}{10^8M_{\odot}} \right)\left[\frac{q}{(1+q)^2} \right]\left(\frac{\sigma}{100 \textrm{km/s}} \right)^{-2}\: \textrm{pc}\,,
 \label{eq:ahard}
\end{equation}
where $q=m_2/m_1\leq1$ is the mass ratio and $\sigma$ is the stellar velocity dispersion. However,
unless mechanisms such  as galaxy rotation or 
merger-induced triaxiality in the galactic gravitational potential help refill the loss cone, or unless other processes that tend to shrink the binary (e.g. Kozai-Lidov resonances due to triple MBH interactions, or gas-induced planetary-like migration) are at play, the binary may stall at separations $a\sim a_{\rm h}$ (``last-parsec problem'').

\subsection{Stalling vs merging binaries}

Unlike previous studies that were conducted with the same semianalytic model, where
MBHs were either assumed to merge at the same time as the host galaxies \citep{2012MNRAS.423.2533B,2014ApJ...794..104S}, or after a suitable ``delay'' time \citep[accounting
for the hardening due to three-body interactions with stars, gas-driven planetary-like migration, and interactions with a third ``intruder'' MBH;][]{2015ApJ...812...72A,2015ApJ...806L...8A,2016PhRvD..93b4003K}, we hereby 
assume that the last-parsec problem is {\it not} solved, and consider three models that bracket the range of possible stalling scenarios for MBH binaries.

In model $A$ we assume that  all MBH binaries stall exactly at the separation
$a_{\rm gw}$ from which GWs would drive them to coalescence in a Hubble time $t_H\approx 13$ Gyr (assuming circular orbits): 
\begin{multline}
 a_{\rm gw}= 7\times 10^{-2}\left(\frac{m_1+m_2}{10^8M_{\odot}} \right)^{3/4}\left[\frac{q}{(1+q)^2} \right]^{1/4}\\\times \left(\frac{t_H}{13 \textrm{Gyr}} \right)^{1/4}\:\textrm{pc}\,.
 \label{eq:agw}
\end{multline}
 
This is intentionally an artificial and pessimistic model, since there is nothing special, physically, about the separation $a_{\rm gw}$. Nonetheless, it will allow us to prove an often under-appreciated point, i.e. the fact that even if all MBH binaries stall,
they may still produce a stochastic GW background detectable by PTAs.

One may, however, argue that model $A$ is actually the most optimistic among all the models where MBH binaries stall, as the stalling radius may be much larger than $a_{\rm gw}$. (Clearly, the stalling radius may not be smaller than $a_{\rm gw}$ otherwise binaries would not stall, but rather coalesce in less than a Hubble time under the effect of GW emission alone). We therefore consider
an even more pessimistic model $B$, where all binaries stall at $a_{\max}=\max(a_{\rm gw},a_{\rm h})$. The hardening radius $a_{\rm h}$ comes about because in this model we are implicitly assuming that the stalling of MBH binaries is due to loss-cone depletion. Moreover, note that Equations (\ref{eq:ahard}) and (\ref{eq:agw}) imply that $a_{\rm h}$ becomes {\it smaller} than $a_{\rm gw}$ for small mass ratios $q\lesssim 10^{-3}$  (i.e. these binaries would merge in less than a Hubble and not stall; see also Fig. 1 in \citealt{2010ApJ...719..851S}). Therefore, in order to be on the conservative side, we take the stalling radius to be the larger between $a_{\rm h}$ and $a_{\rm gw}$. 

To assess what happens when this last assumption is not made, we also consider a model $C$, where we 
place initially all binaries at the  hardening radius
when two galaxies coalesce, and evolve from there {\it under GW emission alone}. 

A few comments are in order here. First, in both models $A$ and $B$, binaries essentially always emit GWs with frequency twice the orbital frequency at the stalling radius (in the source frame).
The signal in model $C$ will instead be dominated by the binaries with $q\lesssim 10^{-3}$, for which $a_{\rm h}<a_{\rm gw}$ and which therefore merge efficiently. 

Second, we note that the stalling radius due to loss-cone depletion might actually be even smaller than $a_{\rm h}$ (by a factor $\sim 5-10$) for comparable mass ratios~\citep{2005LRR.....8....8M}. (Note however that our $a_{\rm h}$, given by Equation \ref{eq:ahard}, agrees to within 20\% with the stalling radius given by Equation 12 in \citealt{2006ApJ...648..976M}, at all mass ratios.) We do not account for this possible effect (which would anyway tend to increase our signal) in neither model $B$ nor $C$, again in order to be on the conservative side. 

Third,  we note that our semi-analytic galaxy-formation model,  despite accounting for the dynamical friction
on the satellite halo/galaxy from the dark matter and the baryon distributions as well as for tidal effects on the satellite, still predicts that a non-negligible
number of unequal-mass galaxy mergers should take place in a Hubble time.
As already mentioned, these systems will in turn form MBH binaries with small mass ratios
$q\lesssim 10^{-3}$, which indeed constitute $\sim 20$\% ($\sim 10$\%) of all the binaries with total mass $10^8 M_{\odot}<M_{\rm tot}< 10^{10} M_{\odot}$ in the light-seed (heavy-seed) case\footnote{Note (M. Volonteri, private communication) that MBH binaries with $q\lesssim 10^{-3}$ in this mass range were also found in a different semi-analytic model, based on \citet{2003ApJ...582..559V}.}. In  model $C$, as discussed above, these binaries 
merge efficiently under GW emission alone, since they are placed at an initial separation $a_{\rm h}< a_{\rm gw}$ when the host galaxies merge. 
Given the potential importance of these systems, the question of whether they are physical (i.e. if it makes sense to place them at the hardening radius after the host galaxies coalesce) deserves further scrutiny. Indeed,
one may wonder whether these binaries will ever reach the hardening radius in the
first place, since (as mentioned above) the smaller MBH may be stripped of the inner parts of its host galaxy early on, thus rendering
dynamical friction from the stellar distribution of the newly formed galaxy inefficient.

However, the stellar bulge of the satellite galaxy only starts getting tidally disrupted when its tidal radius 
$r_{\rm t}$
becomes comparable to its half-light radius $R_{\rm e}$. Since the tidal radius is related to the distance $R$ between the satellite and the centre of the host halo by~\citep{henriques}
\begin{equation}
r_{\rm t}\approx \frac{1}{\sqrt{2}} \frac{\sigma_{\rm sat}}{\sigma_{\rm host}} R\,,
\end{equation}
we obtain that the satellite's bulge starts getting tidally disrupted at a separation 
\begin{equation}
R\approx {\sqrt{2}} \frac{\sigma_{\rm host}}{\sigma_{\rm sat}} R_{\rm e}\,.
\end{equation}
From that separation onwards, the satellite evolution is driven by the dynamical friction of the ``naked'' satellite MBH against the
stellar background of the host. The time needed for the satellite MBH to fall to the centre is therefore~\citep{binneytremaine}
\begin{equation}
t_{\rm DF}\approx \frac{19 {\rm Gyr}}{\ln(1 + M_{\rm h,\star}/M_{\rm bh,s})} \left(\frac{R}{5 {\rm kpc}}\right)^2 \frac{\sigma_{\rm h}}{200 {\rm km/s}}\frac{10^8 M_{\odot}}{M_{\rm bh,s}}\,.
\end{equation}
where the subscripts ``s'' and ``h'' denote the satellite and the host, respectively.
As in \citet{mcwilliams}, we use the results of \citet{oser,nipoti} to assume
\begin{align}
R_{\rm e} =& 2.5\,{\rm kpc} \left(\frac{M_{s,\star}}{10^{11}\,M_\odot}\right)^{0.73}(1+z)^{-1.44}\, \nonumber \\
\sigma_{\rm h}  =& 190\,{\rm km/s} \left(\frac{M_{h,\star}}{10^{11}\,M_\odot}\right)^{0.2}(1+z)^{0.44}\,,\nonumber \\
\sigma_{\rm s}  =& 190\,{\rm km/s} \left(\frac{M_{s,\star}}{10^{11}\,M_\odot}\right)^{0.2}(1+z)^{0.44}\,.
\label{eq:re}
\end{align}
Note that the redshift dependence is valid for $z\lesssim 2$~\citep{oser}.
We then use (for both the host and the satellite) the correlation between black-hole and bulge stellar mass of \citet{kormendy}, lowering the normalisation by a factor $b\sim2$--3 to account for the selection bias (on the resolvability of the MBH sphere of influence) pointed out in \citet{shankar}, to obtain
the final result
\begin{multline}\label{tdf}
t_{\rm DF}\approx 0.38 {\rm Gyr} \\\times b^{1.4} \left(\frac{M_{\rm bh,h}}{10^9M_\odot}\right)^{0.5} \left(\frac{M_{\rm bh,s}}{10^6M_\odot}\right)^{-0.1} (1 + z)^{-2.44}
\\\times \left[1 + 0.07 \ln\left(\frac{b\cdot M_{\rm bh,h}}{10^9M_\odot}\right)-0.08\ln\left(\frac{M_{\rm bh,s}}{10^6M_\odot}\right)\right]^{-1}\,.
\end{multline}
Choosing $M_{\rm bh,s}\approx 10^6 M_\odot$ and $M_{\rm bh,h}\approx 10^9 M_\odot$, $t_{\rm DF}$ becomes comparable to or larger than
the look-back time only for $z\lesssim0.025$ for $b=1$ (the uncorrected relation of \citealt{kormendy}), or for $z\lesssim0.1$ for $b=3$.

However, if we take into account that the selection bias highlighted by \citet{shankar} may not only change the normalisation but might also
make the black-hole -- stellar mass relation steeper, the dynamical friction time may be even longer. If we adopt the intrinsic scaling relation of Equation~6 of \citet{shankar}, we find
\begin{multline}\label{tdf2}
t_{\rm DF}\approx 30 {\rm Gyr} \\\times  \left(\frac{M_{\rm bh,h}}{10^9M_\odot}\right)^{0.3} \left(\frac{M_{\rm bh,s}}{10^6M_\odot}\right)^{-0.46} (1 + z)^{-2.44}
\\\times \left[1 + 0.038 \ln\left(\frac{M_{\rm bh,h}}{10^9M_\odot}\right)-0.075\ln\left(\frac{M_{\rm bh,s}}{10^6M_\odot}\right)\right]^{-1}\,.
\end{multline}
Still, for $M_{\rm bh,s}\approx 10^6 M_\odot$ and $M_{\rm bh,h}\approx 10^9 M_\odot$, even this expression gives a dynamical friction time lower than the lookback time already at $z\approx0.8$. To be on the conservative side, when considering the model where we place all binaries at the hardening radius (model $C$), we 
use Equation~\ref{tdf2} to discard all systems for which $t_{\rm DF}$ is longer than the look-back time (c.f. discussion in Section \ref{sec:res}). Note that since
the redshift dependence is valid for $z\lesssim 2$, we actually replace $z\to \min(z,2)$ in Equation~\ref{tdf2}, in order to avoid artificially short dynamical friction times at high redshift.

Overall, this discussion shows that it makes sense to assume that MBH binaries with $q\lesssim 10^{-3}$ efficiently reach the hardening radius after their host galaxies merge. Nonetheless, even in the absence of such systems, i.e. if all MBH binaries were to stall and not coalesce, we would fall back onto our ``most pessimistic'' model $B$. We will show in the next section that even this model would still be detectable by future PTAs.

\section{Results}
\label{sec:res}

\begin{figure*}
\begin{tabular}{cc}
\epsfig{file=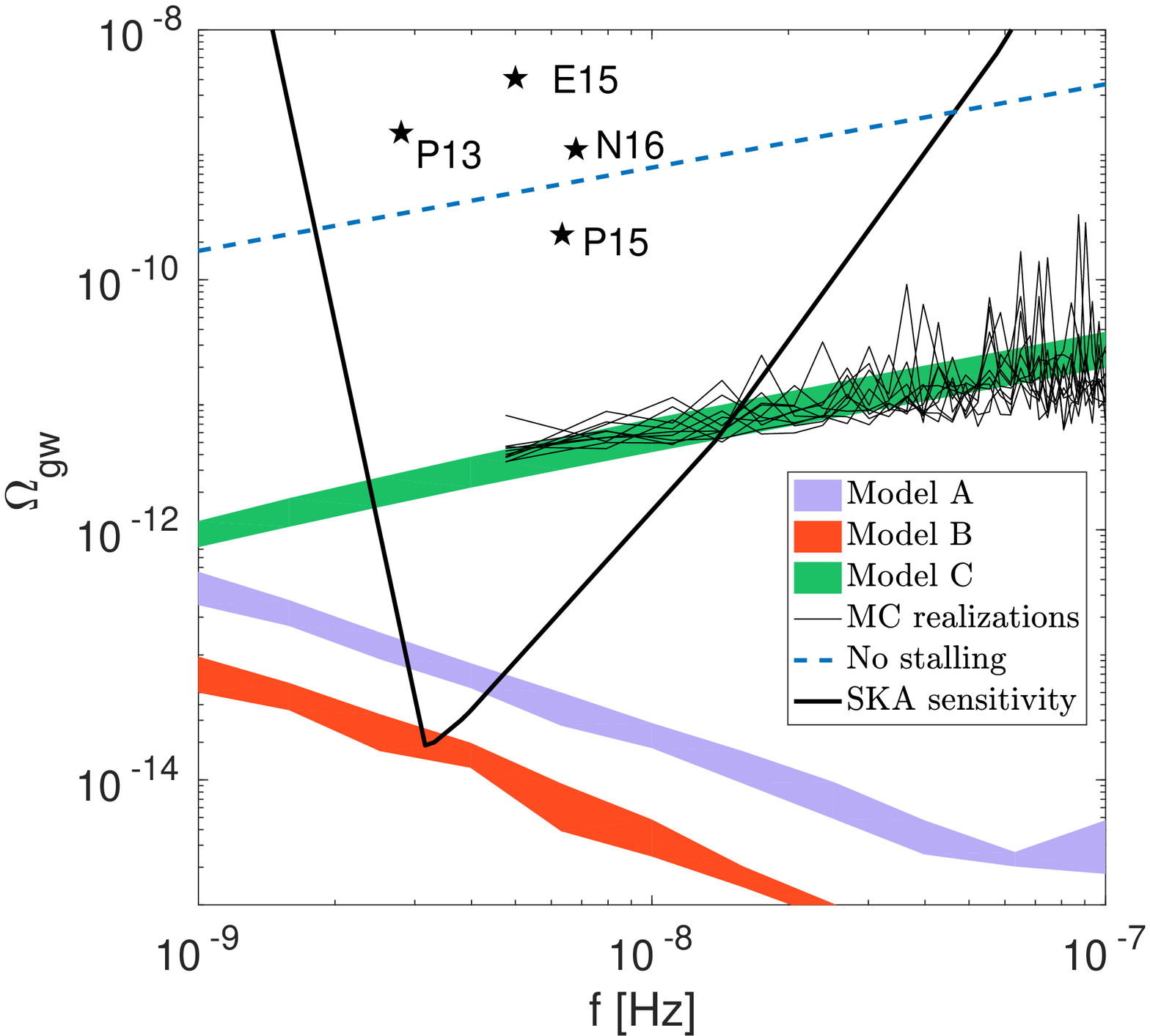, width=.45\textwidth}&
\epsfig{file=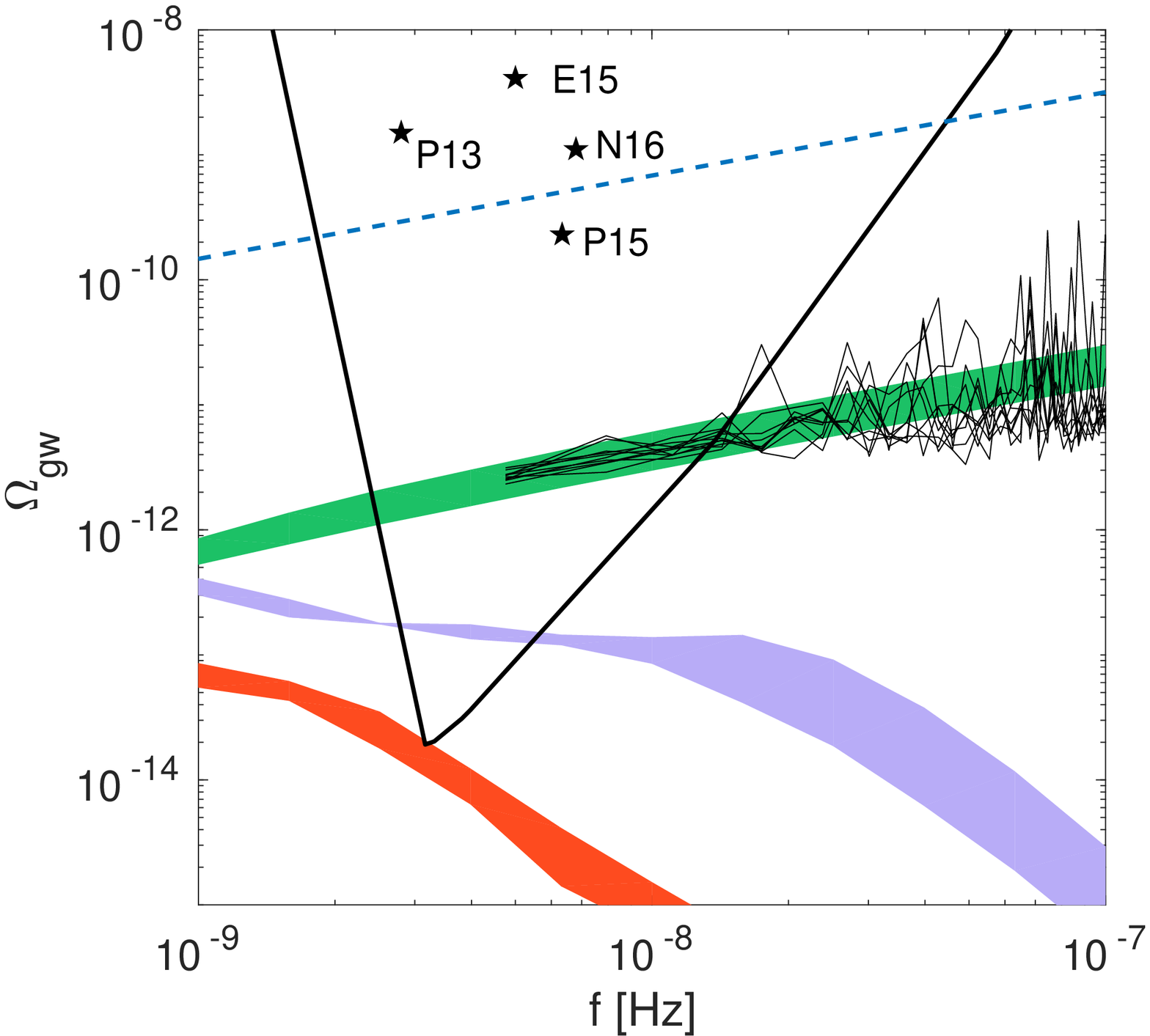, width=.45\textwidth}
\end{tabular}
\caption{The stochastic GW background from MBH binaries in a light-seed model (left panel) and in a heavy-seed one (right panel) in the frequency band of PTA experiments, for models $A$ (all MBH binaries stalling at $a_{\rm gw}$; purple band), $B$ (all MBH binaries stalling at $\max(a_{\rm h},a_{\rm gw})$; red band), and $C$ (all MBH binaries form at $a_{\rm h}$ and are let free to evolve from there under GW emission alone; green band). For model $C$, $10$ different Monte Carlo realisations of the signal are produced by following SVC08 and are shown by thin black lines.
For models $A$ and $B$, similar realisations are very smooth and would be indistinguishable from the purple and red bands (c.f. text for details). For comparison, the background produced if all MBHs binaries merge within a Hubble time is shown by the blue dashed curve.  Also shown is the power-law integrated sensitivity curve for an SKA-based PTA experiment  monitoring $50$ pulsars for $10$ years, by assuming a timing accuracy of $30$ ns (thick black curve). Any power-law spectrum
tangent to this curve gives an SNR $\rho=1$, while any power-law spectrum crossing it yields  an SNR $\rho>1$.
 $95$\% confidence upper limits from current PTA observations are shown as black stars,  and include PPTA  (P15, \citealt{2015Sci...349.1522S}; and P13, \citealt{2013Sci...342..334S}), EPTA \citep[E15;][]{2015MNRAS.453.2576L} and NANOGrav \citep[N16;][]{2016ApJ...821...13A}.}
\label{fig:stalling_omega}
\end{figure*}

We now address the prospects of using future PTA experiments to detect the GW stochastic backgrounds from
the models discussed above, namely model $A$, in which $a_{\rm stall}=a_{\rm gw}$; model $B$, in which $a_{\rm stall}=\max(a_{\rm h},a_{\rm gw})$; and model $C$, where all binaries are assumed to form at a separation $a=a_{\rm h}$, 
and are let evolve under GW emission alone from there (i.e. most of the binaries will not merge by $z=0$, unless
they have low mass ratios $q\lesssim 10^{-3}$, c.f. discussion above). These models are 
represented in Figure \ref{fig:stalling_omega} by respectively purple, red and green bands, in the light-seed (left panel) and heavy-seed (right panel) model for MBHs.

To compute the GW background for models $A$ and $B$, we use two different approximations, as explained in Section \ref{sec:gwb}. In the first,  we 
use Equations~\ref{eq:Omega}, \ref{eq:dEdf1} and \ref{eq:dEdf2}, but we cut off the spectrum given by 
Equation~\ref{eq:dEdf2} outside the frequency interval $[f_{\rm stall}, f_{\rm stall}+\Delta f]$. As explained previously, this method accounts for the orbital evolution
of the binary, which sweeps a finite (albeit small) interval in source-frame frequency from its formation to the present time,
but neglects the change in cosmological redshift during the lifetime of source. In the second approximation, we use Equations~\ref{eq:stalling_spectrum}--\ref{eq:back_stalling}, which account for the varying cosmological redshift during the lifetime of source, but neglect the orbital
evolution of the binary, which is assumed to stall at a fixed separation (i.e. emit at fixed GW frequency in the source frame)
after formation. The difference between these two approximations, which can be thought of as an uncertainty in our 
predictions, is illustrated by the width of the purple and red bands. Note that since the evolution under the influence of GW emission is not significant at the separations $a_{\rm stall}$ considered in these two models, the two approximations yield very similar results.

As for model $C$, to be conservative we neglect the contribution of the binaries for which $a_{\rm h}> a_{\rm gw}$ (which
do not merge in a Hubble time and therefore give a negligible contribution from the signal from this model), and account only
for those with $a_{\rm h}< a_{\rm gw}$. For this subset of binaries, we compute the background by using
 Equations~\ref{eq:Omega}, \ref{eq:dEdf1} and \ref{eq:dEdf2}, but we only consider the spectrum given by 
Equation~\ref{eq:dEdf2} between the initial frequency of the binary, corresponding to the hardening radius, and the final
frequency that it has at $z=0$ (be that finite, if the binary has not yet merged by $z=0$, or formally infinite, if the binary has merged by $z=0$). We have checked that neglecting these cutoffs does not change our results significantly. Note that in this model, unlike in models $A$ and $B$, neglecting the change in cosmological redshift during the evolution of the binary is a very good approximation, since the bulk of the signal comes from binaries that merge. Note also that the 
finite width of the green band in Figure \ref{fig:stalling_omega} accounts for the effect of including or not including binaries for which 
$t_{\rm DF}$ is longer than the look-back time at formation (c.f. discussion at the end of Section \ref{sec:formation}). As can be seen, excluding those systems only makes a small difference.

As far as the spectral shapes of the predictions in Figure \ref{fig:stalling_omega} are concerned, observe that models $A$ and $B$ have somewhat similar behaviour (with energy density decreasing with frequency), though the normalisation is different because of the different stalling radii adopted. As for 
model $C$, the signal is dominated by the subset of merging binaries, hence the spectral dependence
is similar to that of a scenario in which all binaries merge efficiently (i.e. successfully overcome the last-parsec problem), shown by the blue line (and given analytically by Equation~\ref{eq:back_merging}).

We also consider the possibility of a high-frequency turnover/flattening of the GW background,
as a result of the small number of sources that may contribute at high frequencies. To this purpose, we follow SVC08
and generate several Monte-Carlo realisations of the signal as it would be detected, in each frequency bin
of width  $\Delta f=1/T$, by a PTA experiment of duration $T$.

On the one hand, we find that in models $A$ and $B$,  all realisations of the signal are essentially identical to the predictions
obtained by using the expressions in Section \ref{sec:gwb} (which neglect this finite-statistics effect). This is because unlike in the case discussed by SVC08,
in these models there are always many sources contributing to the high-frequency bins of the spectrum. The reason is two-fold.
First,  in models $A$ and $B$ the bulk of the signal comes from lower-mass binaries than in the scenario
considered in SVC08, which assumes that all binaries merge efficiently. Since the MBH mass function
decreases with the MBH mass, lower-mass binary systems are more numerous. Second, in the case of binaries merging efficiently,
a given MBH binary sweeps the entire frequency range of PTAs very quickly, while for a stalling binary the change in frequency (in the detector frame) is much slower and to be ascribed almost entirely to the change in cosmological redshift over the system's lifetime. As a result, simply by using the continuity equation as in SVC08, the expected number of binaries
per frequency bin is much lower for merging binaries than for stalling ones.

On the other hand, for model $C$ the high-frequency part of the signal's spectrum shows features qualitatively similar to those
found in SVC08, with a general flattening or even turnover of the power law, and few pronounced ``spikes''
in few high-frequency bins (several realisations of the signal for an experiment duration $T=10$ yr are shown in black in Figure \ref{fig:stalling_omega}; note that our bins have width $\Delta f=1/T$, therefore the lowest plotted frequency is the midpoint of the first bin, $f=1.5/T$). The resemblance to the results of  SVC08
is not surprising, because in model $C$ the bulk of the signal comes from merging binaries, like in the case of SVC08. (Note however that the number of merging sources in model $C$ is lower than in SVC08, which is reflected by the different normalisation of the green bands and black lines with respect to the blue line.)

Current upper limits from ongoing PTA observations are indicated by black stars in Figure \ref{fig:stalling_omega}. We show results from PPTA  (P15, \citealt{2015Sci...349.1522S}; and P13, \citealt{2013Sci...342..334S}), EPTA \citep[E15;][]{2015MNRAS.453.2576L} and NANOGrav \citep[N16;][]{2016ApJ...821...13A}. Note that the hypothesis of efficient, circular mergers (blue line) is already excluded in our model by the PPTA limits. However, all the other scenarios (models $A$, $B$ and $C$) considered in this paper are still below the observed upper limits, though they may be tested with more sensitive experiments. 

In order to estimate the detection prospects with future PTAs, we consider an SKA-like experiment  monitoring $50$ pulsars with $30$ ns timing accuracy for $T=10$ yr, and calculate a power-law integrated sensitivity curve \citep[by using the procedure in][thick black line]{2013PhRvD..88l4032T}. By construction, any power-law spectrum  tangent to this
sensitivity curve has a signal-to-noise ratio (SNR) of $\rho=1$,
and a power-law spectrum that crosses it would have $\rho>1$. Therefore, 
as can be seen from Figure \ref{fig:stalling_omega},
models  $A$ and $C$ would be easily detectable by such an SKA-based PTA, and even
 the most pessimistic scenario (model $B$) may be marginally detectable, as we discuss in detail below.

We note also that the results from the heavy- and light-seed models are very similar to one another in the range of frequencies relevant for present and future PTAs. Indeed, the main difference between the two models is 
the modest bump at $\sim 20$ nHz that can be seen in the heavy-seed model. This feature is due to
binaries with chirp mass $\sim 10^5 M_\odot$, i.e. stalling binaries of MBHs that have not evolved much from their initial seed masses. 
(``Seed binaries'' are also present in the light-seed model, but because of their lower masses they radiate at higher frequencies).
In the following we show only the results for the light-seed model.

The power-law integrated sensitivity curve shown in Figure \ref{fig:stalling_omega}  is computed in the weak-signal limit, i.e. under the assumption that the GW background is sub-dominant with respect to the intrinsic white-noise component \citep{2015PhRvD..91d4048C}. The white-noise power spectrum is $\mathcal{P}_N=2\sigma^2\Delta t$ where $\sigma$ is the pulsar timing accuracy and $\Delta t$ is the cadence of the pulsar measurement. For an SKA-based PTAs,
reasonable typical values may be $\sigma=30$ ns and $\Delta t=\,$yr$/20$, so the background signal 
produced by our models would be in the intermediate regime: It dominates the noise at low frequencies, but is sub-dominant at high frequencies. 
In this regime, we therefore have to use 
the general expression for the SNR \citep{2009PhRvD..79h4030A,2013CQGra..30v4015S,2015PhRvD..91d4048C}:
\begin{equation}
 \langle \rho \rangle = \left(\sum_{IJ} \chi^2_{IJ} \right)^{1/2}\left(2T\int_{f_{\rm L}}^{f_{\rm H}}df\frac{\mathcal{P}^2_g(f)}{\left(\mathcal{P}_{\rm g}(f)+2\sigma^2\Delta t \right)^2} \right)^{1/2}\,,
 \label{eq:SNR}
\end{equation}
where $T$ is the total observation time, $\chi^2_{IJ}$ is the Hellings and Downs coefficient for pulsars $I$ and $J$ \citep{1983ApJ...265L..39H} and $\mathcal{P}_{\rm g}(f)$ is the power spectrum of the signal. Note that the lower limit of the integral $f_{\rm L}=1/T$ is set by the total observation time $T$. 

\begin{figure}
\centering
\epsfig{file=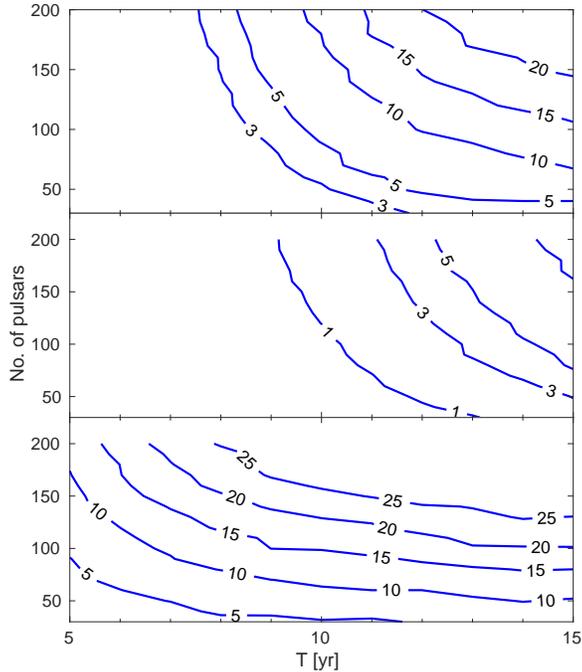, height=10cm}
\caption{SNR as a function of observation time $T$ and number of pulsars (assuming an SKA-based PTA experiment with a timing accuracy of $30$ ns), in the light-seed model. The SNR is computed by using Equation \ref{eq:SNR} for the different scenarios discussed in the text, namely model $A$ (top panel), $B$ (middle panel) and $C$ (bottom panel). Note that even in the most pessimistic case, shown in the middle panel, the signal from stalling binaries can be observed with SNR $\rho=5$ after $15$ years of monitoring $100$ pulsars.}
\label{fig:snr_contours}
\end{figure}

In Figure \ref{fig:snr_contours}, we show the SNR by assuming an SKA-based PTA with $\sigma=30$ ns and $\Delta t=\,$yr$/20$,
as a function of number of pulsars (which we distribute isotropically in the sky) and observation time. The upper and middle panels correspond to models $A$
and $B$, respectively. In both cases, the SNR is very sensitive to the total observation time $T$ because of the steep frequency dependence of the signal (as seen in Figure \ref{fig:stalling_omega}). As a result of this frequency dependence, only a small frequency range around $f_{\rm L}$ contributes to the integral in Equation \ref{eq:SNR}. We find that the signal can be detected with an SKA-based PTA experiment in both cases: for binaries stalling at $a_{\rm gw}$ (model $A$) or $\max(a_{\rm gw},a_{\rm h})$ (model $B$) an SNR of $\rho=5$ ($\rho=3$) can be obtained with $\sim 100$ ($\sim 50$) pulsars, and $10$ or $15$ years of observations respectively for models $A$ and $B$. 
Detection prospects are even better for our model $C$ (bottom panel in Figure \ref{fig:snr_contours}): $\rho=5$ can be obtained after only $5$ years of monitoring $\sim 70$ pulsars. 

Current PTAs (PPTA, EPTA and NANOGrav) have worse timing accuracies than those assumed above, but have already been gathering data for several years and have built long timing baselines. Assuming a timing accuracy of only $\sigma=250$ ns, current experiments will need to monitor $\sim 70$ pulsars for a duration of $\sim 15$ years in order to detect the signal from model \emph{C} with an SNR of $\rho=5$. Taking into account the data already gathered by the different PTAs, this detection might thus be not too far in the future. Detecting the signal from stalling binaries will be more challenging: in the case of models \emph{A} and \emph{B}, a detection with $\rho=5$ will require monitoring $\sim 100$ pulsars (all with timing accuracy of $\sigma=250$ ns) for a duration of $20$ and $30$ years, respectively.

\begin{figure}
\centering
\epsfig{file=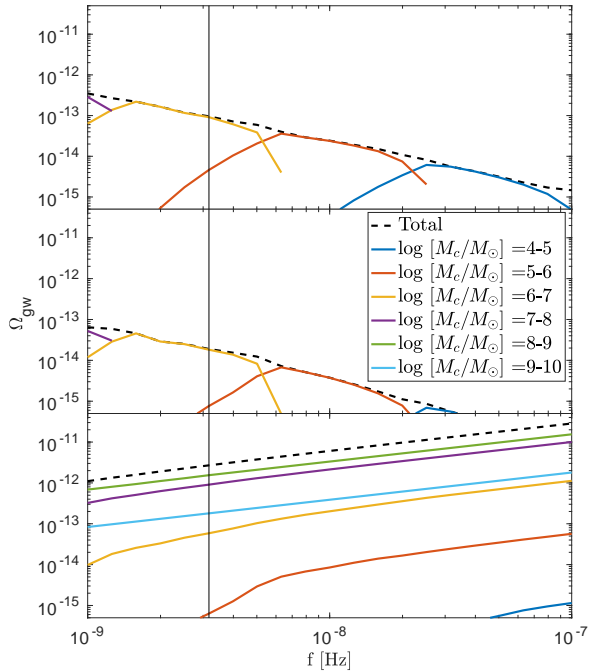, height=10cm}
\caption{GW energy density from MBH binaries in the light-seed model and for the different scenarios
discussed in the text, namely model $A$ (top panel), model $B$ (middle panel) and model $C$ (bottom panel). Different chirp mass ranges are indicated in the legend, and the total signal is shown by the black dashed line. The vertical black line indicates $f_{\rm L}=1/T$, with $T=10$ yr.}
\label{fig:hc_mass_freq}
\end{figure}

Let us now explore the range of masses of the MBH binaries that contribute to the background signal. In Figure \ref{fig:hc_mass_freq}, we show the contributions to the energy density for different ranges of the binary's chirp mass. The vertical line denotes the limiting frequency $f_{\rm L}$  corresponding to $10$ years of observation. For binaries that stall at $a_{\rm gw}$ (model $A$) or $\max(a_{\rm gw},a_{\rm h})$ (model $B$), which are represented respectively in the upper and middle panels, the signal at $f_{\rm L}$ is dominated by systems in the chirp-mass range $10^{6}-10^{7}M_{\odot}$, with a smaller contribution from the $10^{5}-10^{6}M_{\odot}$ range. This is exactly the range of masses targeted by LISA. In other words, our results suggest that these MBH binaries will be observed either by LISA if they merge efficiently (i.e. within a Hubble time; c.f. \citealt{2016PhRvD..93b4003K}), or by SKA-based PTAs if they stall. In the case of model $C$ (bottom panel), the mass distribution is instead quite different, with the signal being dominated by systems in the range $M_c=10^{7}-10^{9}M_{\odot}$ at all frequencies. This is
very similar to the masses of the binaries that contribute to the PTA GW background under the hypothesis that the
final-parsec problem is efficiently solved (blue lines in Figure \ref{fig:stalling_omega}), c.f. SVC08.
Clearly, this resemblance comes about because, as already mentioned, the signal in model $C$
is dominated by a sub-population of binaries 
for which the final-parsec problem is not relevant, because they have 
$a_{\rm h}< a_{\rm gw}$, and thus coalesce in less than a Hubble time under the effect of GW emission alone.

\begin{figure}
\centering
\epsfig{file=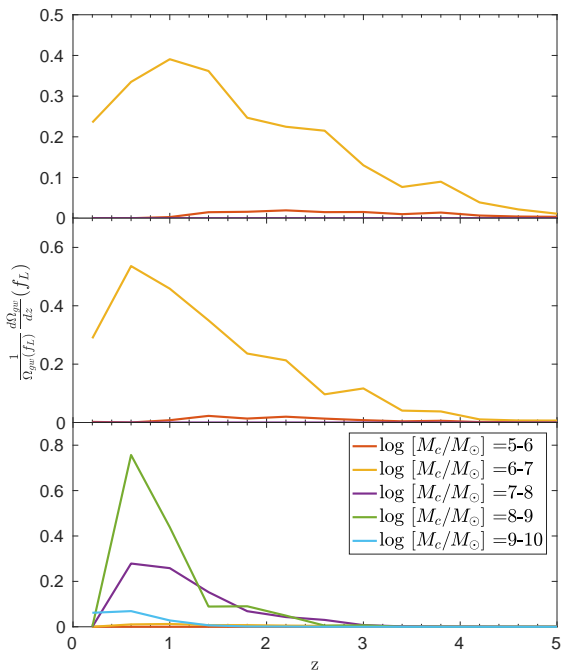, height=10cm}
\caption{Energy density distribution at $f_{\rm L}=1/T$ (with $T=10$ yr) in the light-seed model and for the different scenarios
discussed in the text, namely model $A$ (top panel), model $B$ (middle panel) and model $C$ (bottom panel). Different chirp-mass ranges are indicated in the legend.}
\label{fig:hc_mass_z}
\end{figure}

In Figure \ref{fig:hc_mass_z}, we also show the GW energy density distribution $d\Omega_{\rm gw}(f_{\rm L})/dz/\Omega_{\rm gw}(f_{\rm L})$ at $f_{\rm L}=1/(10$ yr$)$ in different chirp-mass and redshift ranges. Note that in our model $C$ (bottom panel) the distribution is sharply peaked at $z\sim 1$, whereas in the case of stalling binaries (i.e. models $A$ and $B$; upper and middle panels) the distribution is significantly broader.

Finally, we have investigated whether the sub-population of merging binaries with $q\lesssim 10^{-3}$ predicted by model $C$ contains systems that would be observable by LISA with significant detection rates.
Indeed, binaries with $q\lesssim 10^{-3}$ may in principle be detectable by LISA as IMRIs.
We find that the light-seed model predicts that about 1 such event would be detectable every 2 years, if we adopt the latest LISA sensitivity curve described in \citet{2017arXiv170200786A}. (By comparison, the LISA mission will last at least 4 years, with a possible extension to up to 10 years.) The detectable events typically have total (source-frame) masses between a few $10^5 M_\odot$ and a few $10^7 M_\odot$, mass ratios $q$ between a few $10^{-4}$ and a few $10^{-3}$, redshift distribution peaked around $z=2-3$ and extending up to $z\sim 5$, and typical SNR $\rho\sim50-200$.
Therefore, they are formed by a MBH with mass corresponding the LISA frequency range, and a second black hole of mass $\sim 10^3 -10^4 M_\odot$, which has not accreted much during its previous history and whose mass is therefore close to the seed mass 
$M_{\rm seed}\sim 200 M_\odot$. 
Conversely, in our heavy-seed scenario the seed mass is $M_{\rm seed}\sim 10^5 M_{\odot}$, so systems with mass ratio
$q\lesssim 10^{-3}$ and including a seed black hole would have a total mass too large to emit a strong signal in the LISA band. In fact, we have verified that in the heavy-seed scenario we only obtain $\sim 0.07$ IMRIs per year that are detectable by LISA.

\section{Summary}
\label{sec:dis}

Recent upper limits from PTA experiments (and especially PPTA, \citealt{2015Sci...349.1522S}) are starting to be in tension with some of the current theoretical estimates for the merger rate of MBH binaries. This finding can be interpreted as due to the
influence of the environment (gas and/or stars) on MBH binaries while they are in the PTA band~\citep{2011MNRAS.411.1467K,2014MNRAS.442...56R,vallisneri,2015PhRvD..91h4055S,2017arXiv170202180K}; to a possible
residual eccentricity~\citep{vallisneri}, resulting for instance from triple MBH interactions; or to a wrong normalisation
of the theoretical predictions due to selection biases in the observations against which they are calibrated~\citep{2016MNRAS.463L...6S}.
However, a much more worrisome possibility (not only for PTA experiments but also for LISA) 
is that MBH binaries may be unable to evolve past the hardening radius $a_{\rm h}\sim$ pc (last-parsec problem)
and stall there.
In this paper, we have used a state-of-the-art semi-analytic galaxy-formation model
including MBHs (isolated and in binaries) to study the stochastic GW background from populations of stalling binaries and its detection prospects with future PTAs. We have presented two major findings:
\begin{itemize}
 \item Even in the least favorable scenarios, the GW background produced by stalling MBH binaries might be observable with the next generation of PTAs \citep[see also][]{vallisneri}. Specifically, if MBH binaries stall at the separation $a_{\rm gw}$ 
from which they would still need a Hubble time to merge, the resulting background can be observed with a SNR of $\rho=5$ ($\rho=3$) with an SKA-like experiment that monitors $\sim 100$ ($\sim 50$) pulsars after $10$ years of observations. Even in the most pessimistic case, in which the binaries stall at $\max(a_{\rm gw},a_{\rm h})$, the same SNR of $\rho=5$ ($\rho=3$) can be achieved with $\sim 100$ ($\sim 50$) pulsars after $15$ years (Figure \ref{fig:snr_contours}). This signal is dominated by binaries in the chirp mass range $10^{6}-10^{7}M_{\odot}$. Therefore, according to our results, binaries in this mass range will be detected either by LISA if they merge, or by an SKA-based PTA experiment if they stall. Observations with the timing accuracies of current PTAs will require a timing baseline of $\sim20-30$ years with $\sim 100$ pulsars.
 \item Our model predicts the existence of a sub-population of MBH binaries with low mass ratios $q\lesssim 10^{-3}$
and  hardening radii sufficiently small to allow these binaries to merge
 within a Hubble time under the effect of GW emission alone 
($a_{\rm h}<a_{\rm gw}$). This sub-population of binaries produces a strong GW background signal that will be easily observable by the next generation of PTAs, requiring only $5$ years of observations with $70$ pulsars at SKA sensitivity to obtain an SNR of $\rho=5$. The timing accuracies achievable with current PTAs will likely require about $15$ years of observing time for the same number of pulsars, but in view of the data already gathered by these experiments, the time to detection might actually be shorter.
We have also shown that the formation of these binaries 
is not an artifact of the simplified prescriptions used in the
semi-analytic model to account for the orbital evolution of merging galaxies and MBH binaries.
Moreover, this sub-population may yield a few detectable events for the LISA mission, if MBHs form from the remnants of popIII stars at high redshifts.
\end{itemize}

\section*{Acknowledgements}
We thank Alberto Sesana, Joe Silk and Marta Volonteri for many invaluable insights and discussions about the issues presented in this paper.
This work has been financially supported by the Programme National Hautes Energies (PNHE) funded by CNRS/INSU-IN2P3, CEA and CNES, France. The work of ID has been done within the Labex ILP (reference ANR-10-LABX-63) part of the Idex SUPER, and received financial state aid managed by the Agence Nationale de la Recherche, as part of the programme Investissements d'avenir under the reference ANR-11-IDEX-0004-02. EB acknowledges support from the H2020-MSCA-RISE-2015 Grant No. StronGrHEP-690904
and from the APACHE grant (ANR-16-CE31-0001) of the French Agence Nationale de la Recherche.
This work has made use of the Horizon Cluster, hosted by the Institut d'Astrophysique de Paris. We thank Stephane Rouberol for running smoothly this cluster for us.

\bibliographystyle{mn2e}
\bibliography{stalling}
\label{lastpage}

\end{document}